\documentstyle[12pt]{article}
\begin{document}
\title{{\bf Entropy for
extremal Reissner- Nordstr\o m black holes}}
\author{Amit Ghosh\thanks{e-mail amit@saha.ernet.in}\\
Saha Institute of Nuclear Physics\\
Block AF, Bidhannagar\\
Calcutta 700 064, INDIA\\and\\
P. Mitra\thanks{e-mail mitra@tsmi19.sissa.it, mitra@saha.ernet.in}\\
Scuola Internazionale Superiore di Studi Avanzati\\
Via Beirut 2-4\\
34014 Trieste, ITALY}
\date{November 1994\\hep-th/9411128\\SISSA-172/94/EP}
\maketitle
\abstract{
We reinterpret recent work on extremal Reissner- Nordstr\o m black
holes to argue that the mass is a
measure of the gravitational entropy.  We also find that
the  entropy of scalar fields in this
background has a stronger divergence than usual.
}
\newpage

It has long been known that the area of the horizon of a black
hole  can  be  interpreted as an entropy \cite{Bek} and satisfies
all  the  thermodynamical laws. This is not fully understood
in terms of the usual formulation of entropy as a measure of
the number of states available, but the na\"\i ve  Lagrangian  path
integral does lead to  a  partition  function  from  which the area
formula for entropy  can  be  obtained  \cite{GH}  by  neglecting
quantum fluctuations.

That formula is supposed to describe the gravitational entropy
corresponding to a black hole.
There have also been investigations on the entropy of quantum
fields  in  black  hole  backgrounds \cite{'tHooft,Uglum,GM}. The values
thus obtained are additional contributions   to  the
entropy of the black hole -field system.
These calculations have produced
divergences,  but the area of the horizon has appeared as a factor.
This has been interpreted to mean that the gravitational
constant gets renormalized in the presence of the quantum fields \cite{Uglum}.

Of special interest is the case of extremal black holes which seem
to have peculiarities not always present in the corresponding
nonextremal cases \cite{Pres}.
For extremal Reissner- Nordstr\o m black holes,
the temperature is zero, but
the {\it area} is nonzero, though the entropy  of a scalar field vanishes.
For extremal dilatonic black holes, where
the temperature is nonzero, the area vanishes,
but the entropy of the scalar field does {\it not} \cite{GM}.

Recently some startling observations have been made \cite{HHR,Teit}
about extremal black holes.
Topological arguments have been presented in the context of Euclidean quantum
gravity to suggest that extremal Reissner- Nordstr\o m black holes
have zero gravitational entropy in spite of
the nonvanishing area and \cite{HHR} {\it no definite temperature.}

What has actually been shown \cite{HHR,Teit}
is that the classical {\it action} for
the Euclidean black hole configurations vanishes, and this has been
argued to lead semiclassically to a vanishing {\it entropy}. In this note
we shall a) study this connection more closely and b)
investigate what happens to the entropy of a scalar
field in the background of such a black hole.

The metric of the Reissner- Nordstr\o m spacetime is given by
\begin{equation}
ds^2=-(1-{2M\over r}+{Q^2\over r^2})dt^2+ (1-{2M\over r}+{Q^2\over r^2})
^{-1}dr^2 +r^2d\Omega^2
\end{equation}
in general, with $M$ and $Q$ denoting the mass and the charge
respectively. There are apparent singularities at
\begin{equation}
r_\pm=M\pm\sqrt{M^2-Q^2}
\end{equation}
provided $M\ge Q$. Cosmic censorship dictates that this
inequality holds and then there is a horizon at $r_+$.
The limiting case when $Q=M$ and $r_+=r_-$ is referred to as
the extremal case.

The formula for the Hawking temperature of such a
black hole is
\begin{equation}
T={r_+-r_-\over 4\pi r_+^2}.
\end{equation}
Clearly this expression vanishes in the extremal case.

The first law of thermodynamics can be written as
\begin{equation}
\tilde TdS=dM-\tilde\Phi dQ,
\end{equation}
where we have inserted tildes to indicate that these quantities
may differ from the usual ones. $\tilde\Phi$ is the analogue
of the chemical potential for the charge and is essentially the
electrostatic potential. $\tilde \Phi$ and $\tilde T$ may differ
from the corresponding expressions in the older literature
because of several reasons: (i) an alternative way of introducing
entropy \cite{GM2}, (ii) quantum corrections \cite{York},
and (iii) the newly discovered properties of extremal black
holes \cite{HHR,Teit} which cannot be continuously deformed
into nonextremal black holes and which, moreover, can be in equilibrium
at different temperatures. We shall be interested
only in the extremal case, so we set $Q=M$ rightaway. Then
\begin{equation}
\tilde TdS=(1-\tilde\Phi)dM.\label{FL}
\end{equation}

We now appeal to the vanishing Euclidean action \cite{Teit}.
The partition  function can be approximately written as
\begin{equation}
Z=e^{-I},
\end{equation}
with $I=0$. Now note that this is the grand canonical partition
function \cite{GH} because the charge
$Q$ is  variable. It is not an {\it independent}
variable if we consider black holes which remain extremal, but it has to
change in processes where the energy $M$ changes. So we write the
usual formula
\begin{equation}
Z=e^{-W/\tilde T},
\end{equation}
with
\begin{equation}
W=M-\tilde TS-\tilde\Phi Q.
\end{equation}
Noting that $W$ has to vanish, we arrive at the relation
\begin{equation}
\tilde TS=(1-\tilde\Phi)M.
\end{equation}
Comparing with (\ref{FL}), and assuming the factors to
be nonvanishing, we conclude that
\begin{equation}
{dS\over S}={dM\over M},
\end{equation}
{\it i.e.,}
\begin{equation}
S=kM, \label{k}
\end{equation}
where $k$ is an undetermined constant. If $k$ vanishes,
we are back to a vanishing entropy, but (\ref{k}) is more generally valid.
It is crucial here to abandon the usual expectation $\Phi=1$
obtained by assuming continuity of the transition from non-extremal to
extremal black holes. As has been argued in
\cite{HHR}, the temperature cannot be fixed.
The potential too cannot be fixed, but it is related
to the temperature by the formula
\begin{equation}
\tilde\Phi=1-k\tilde T.
\end{equation}
The only definite result is that the thermodynamical
entropy, in contradistinction to the action, is
given by the mass of the extremal black hole upto an unimportant
constant.

We shall now study scalar matter in this spacetime. We   employ   the
brick-wall boundary condition \cite{'tHooft}. Then the
wave function is cut off just outside the horizon. Mathematically,
\begin{equation}
\varphi(x)=0\qquad {\rm at}\;r=r_++\epsilon
\end{equation}
where   $\epsilon$  is a small, positive, quantity and signifies
an ultraviolet cut-off. There is also an infrared cut-off
\begin{equation}
\varphi(x)=0\qquad {\rm at}\;r=L
\end{equation}
with the box size  $L>>r_+$.

The  wave  equation  for  a scalar  is
\begin{equation}
{1\over\sqrt{-g}}\partial_{\mu}(\sqrt{-g}g^{\mu\nu}\partial_{\nu}\varphi)
-m^2\varphi=0.
\end{equation}
A solution of the form
\begin{equation}
\varphi=e^{-iEt}f_{El}Y_{lm_l}
\end{equation}
satisfies the radial equation
\begin{eqnarray}
(1-{2M\over r}+{Q^2\over r^2})^{-1}E^2f_{El}&+&{1\over r^2}{\partial\over
\partial r}[(r^2-2Mr+Q^2){\partial f_{El}\over\partial r}]\nonumber\\
&-&[{l(l+1)\over r^2}+m^2]f_{El}=0.
\end{eqnarray}
An $r$- dependent radial wave number can be introduced from  this
equation by
\begin{equation}
k^2(r,  l,  E)=  (1-{2M\over  r}+{Q^2\over r^2})^{-1}
[(1-{2M\over r}+{Q^2\over r^2})^{-1} E^2 -
{l(l+1)\over r^2} -m^2].
\end{equation}
Only such values of $E$ are to be considered here as make the  above
expression  nonnegative. The values are further restricted by
the semiclassical quantization condition
\begin{equation}
n_r\pi=\int_{r_++\epsilon}^L~dr~k(r, l, E),
\end{equation}
where $n_r$ has to be a nonnegative integer.

To find the free energy $F$ at inverse temperature $\beta$
one has to sum over states with all possible single- particle
combinations:
\begin{eqnarray}
\beta F&=&\sum_{n_r, l, m_l}\log(1-e^{-\beta E})\nonumber \\
&\approx  &  \int  dl~(2l+1)\int  dn_r\log   (1-e^{-\beta   E})
\nonumber\\
&=&-\int  dl~(2l+1)\int d(\beta E)~(e^{\beta E} -1)^{-1} n_r \nonumber\\
&=& -{\beta\over\pi}\int  dl~(2l+1)
\int dE~(e^{\beta E} -1)^{-1}\int_{r_++\epsilon}^L
dr~(1-{2M\over r}+{Q^2\over r^2})^{-1}\nonumber\\
&& \sqrt{E^2-(1-{2M\over r}+{Q^2\over r^2})({l(l+1)\over r^2}+m^2)} \nonumber\\
&=& -{2\beta\over 3\pi}\int_{r_++\epsilon}^L dr~ (1-{2M\over r}+
{Q^2\over r^2})^{-2}
r^2\nonumber\\&& \int dE~(e^{\beta E} -1)^{-1}
[E^2-(1-{2M\over r}+{Q^2\over r^2})m^2]^{3/2}.
\end{eqnarray}
Here  the  limits  of  integration  for  $l, E$ are such that the
arguments  of  the  square  roots  are   nonnegative.   The   $l$
integration  is  straightforward  and has been explicitly carried
out. The $E$ integral can be evaluated only approximately.

The contribution to the $r$ integral from  large  values  of  $r$
yields   the  expression  for  the  free  energy  valid  in  flat
spacetime:
\begin{equation}
F_0=-{2\over 9\pi}L^3\int_m^\infty dE{(E^2-m^2)^{3/2} \over
e^{\beta E} -1}.
\end{equation}
We ignore this part \cite{GH,'tHooft}.
The contribution of the hole  is  singular  in  the  limit
$\epsilon\to 0$. The leading singularity for a nonextremal
black hole is linear:
\begin{equation}
F_{non-ex}\approx -{2\pi^3\over  45\epsilon}(1-{r_-\over r_+})^{-2}
({r_+\over\beta})^4,
\end{equation}
where  the lower limit of the $E$ integral has been approximately
set equal to zero. If  the  proper  value  is  taken,  there  are
corrections  involving  $m^2\beta^2$  which will be ignored here.
This  result  reduces  to   the   formula \cite{'tHooft}  for   the
Schwarzschild  black hole and more generally yields an entropy
proportional to the area if $\beta$ is replaced by the reciprocal
of the Hawking temperature and the na\"\i ve cutoff $\epsilon$
replaced by an invariant one as in \cite{'tHooft}.

The result changes if the black hole is extremal:
\begin{equation}
F_{ex}\approx -{2\pi^3r_+^2\over  135\epsilon^3}
({r_+\over\beta})^4.
\end{equation}

The contribution to the entropy due to the
presence of the black hole can be obtained from the formula
\begin{equation}
S=\beta^2 {\partial F\over\partial\beta}.
\end{equation}
This gives
\begin{equation}
S_{ex}=  {8\pi^3\over  135}({r_+\over\beta})^3({r_+
\over\epsilon})^3.\label{S}
\end{equation}

If the usual vanishing expression for the temperature is accepted,
the expression (\ref{S}) for the entropy also
vanishes and this has been the understanding about this
entropy until  recently. However, the temperature may
be nonvanishing, because of quantum corrections and because of
the new observation \cite{HHR} that the Euclidean
solution can be identified with an {\it arbitrary} period $\beta$.
For a general $\beta$ (\ref{S}) is nonzero and in fact cubically
divergent, whereas nonextremal black holes have only a linear
divergence.

Thus we have obtained a rather surprising result
for the entropy of a scalar field in the background of an
extremal Reissner- Nordstr\o m black hole. Whereas
the gravitational entropy is proportional not to $M^2$ but to zero or $M$,
depending on whether one prefers
the original version of the recent work of \cite{HHR,Teit} or
our reinterpretation (\ref{k}) of that work given above,
the entropy of the scalar field is  even
{\it more} singular than in the nonextremal case. The background
of an extreme dilatonic black hole also gave a nonzero result
when a zero was expected,
but there the leading linear singularity did drop out and only a logarithmic
term remained. The Reissner- Nordstr\o m result is therefore
even more  surprising.

\bigskip

\noindent{\bf Acknowledgment}: PM wishes to thank SISSA for its
hospitality.

\end{document}